\documentclass[fleqn]{2023SCGE}
\begin{document}

\ensubject{subject}

\ArticleType{Article}
\SpecialTopic{SPECIAL TOPIC: }

\title{Cascading nuclear excitation of \(^{235}\text{U}\) via inelastic electron scattering in laser-irradiated clusters}{Cascading nuclear excitation of \(^{235}\text{U}\) via inelastic electron scattering in laser-irradiated clusters}


\author[1]{Qiong Xiao\textsuperscript{\dag}}{}
\author[1]{Yang-Yang Xu\textsuperscript{\dag}}{}
\author[1]{Jun-Hao Cheng}{}
\author[2]{\\Liang-Qi Zhang}{}
\author[1]{Wen-Yu Zhang}{}
\author[2]{Yong-Sheng Huang}{}
\author[1]{Tong-Pu Yu\textsuperscript{*}}{}%

\footnotetext[1]{\textsuperscript{\dag} These authors contributed equally to this work.}
\footnotetext[2]{\textsuperscript{*} Corresponding author (email:tongpu@nudt.edu.cn).}

\AuthorMark{Q. Xiao}

\AuthorCitation{Q. Xiao, Y. Y. Xu, J. H. Cheng, et al}

\address[1]{College of Science, National University of Defense Technology, 410073 Changsha, People's Republic of China}
\address[2]{School of Science, Shenzhen Campus of Sun Yat-sen University, 518107 Shenzhen, People's Republic of China}


\abstract{Nuclear excitation induced by lasers holds broad application prospects in precision metrology and nuclear energy, such as nuclear batteries and nuclear clocks. In the present work, the nuclear excitation via inelastic electron scattering (NEIES) mechanism in \(^{235}\text{U}\) is investigated by combining the Dirac-Hartree-Fock-Slater method for theoretical calculations with 2D3V particle-in-cell (PIC) simulations for numerical modeling. The excitation cross-sections of \(^{235}\text{U}\) from the ground state to excited states is evaluated and an efficient indirect excitation scheme for generating \(^{235}\text{U}\) isomers is proposed by first exciting the nuclei from ground-state to high-energy excited states and then decays to the isomeric state. The calculations show that laser-tuned multi-channel NEIES can boost isomeric state accumulation efficiency by eleven orders of magnitude over direct excitation.  Additionally, PIC simulations reveal that laser polarization significantly alters high-energy electron distributions. At identical intensity, p-polarized lasers generate seven times more electrons with energies sufficient for exciting $^{235}\text{U}$ to the second excited state (and beyond to the fourth state) compared to s-polarization. This work offers a practical and feasible way for efficient isomeric state generation, supporting experimental tests of laser-driven nuclear excitation in sub-TW lasers.}



\maketitle

\begin{multicols}{2}

\section{Introduction}

The rapid development of laser technology has made laser an indispensable and powerful tool in many scientific and technological fields, with its study focus having shifted from initially achieving stable coherent radiation \cite{MAIMAN1960} to, in modern times, attaining extremely high laser intensities--up to \(10^{23}\ \text{W/cm}^2\)--and ultra-short pulse durations in the femtosecond to attosecond range \cite{STRICKLAND1985447, Yoon21, Xu2024}. This technological advancement in lasers has expanded the possibilities for laser applications in multiple fields and opened new domains in nuclear physics and astrophysics \cite{doi10.1126/science.1080552}. Traditionally, the fundamental research in nuclear physics has relied mainly on methods such as radioactive sources, linac accelerators, and reactors to explore nuclear properties and nuclear structure \cite{CPC-2021-0034,CPC-2020-0033,NUBASE2020}. These facilities are normally large, costly, and inaccessible to normal users, limiting their wide applications. Fortunately, the unique ability of lasers, e.g., generating highly localized and temporally precise electromagnetic fields and extremely high laser intensities, offers exciting new opportunities for studying nuclear phenomena and nuclear energy, etc \cite{Gales_2018, 10.1063/1.5093535}.

In recent years, significant progress has been achieved in laser-nucleus interactions, garnering widespread attention around the world \cite{PhysRevLett.124.212505,PhysRevLett.119.202501,PhysRevLett.133.152503,PhysRevLett.133.132501,PhysRevC.110.064326,PhysRevLett.128.162501,PhysRevC.103.014313}. Theoretically, researches on laser-nucleus interactions have advanced through frameworks such as the time-dependent Schr\"{o}dinger equation and Fermi's Golden Rule, demonstrating the feasibility of laser manipulation of nuclear behavior in processes like laser-assisted charged particle emission and low-energy nuclear excitation \cite{cjh2025,10.1063/5.0212163,Xiao2024,Zou_2024,CHENG2024138322,PhysRevC.105.024312,PhysRevC.106.064604,PhysRevC.110.064621,PhysRevC.106.044604,PhysRevLett.124.242501}. Experimentally, researchers have achieved laser-induced nuclear excitation such as $^{83}$Kr \cite{PhysRevLett.128.052501,pnas2413221121}, $^{45}$Sc \cite{Shvydko2023}, $^{93}$Mo \cite{chiara2018isomer,PhysRevLett.128.242502}, and $^{229}$Th \cite{Kraemer2023,PhysRevLett.132.182501,PhysRevLett.133.013201,Zhang2024,Zhang20242} in recent years.  The interdisciplinary nature of these studies combines aspects of laser physics, nuclear structure, and quantum dynamics, enabling unprecedented manipulation of nuclear systems under extreme electromagnetic fields or electron-rich environments. These researches hold opportunities for many potential applications in precision metrology, laboratory astrophysics, and nuclear energy \cite{PARIKH2013225,Misch_2021,10.3389/fphy.2024.1503516}. 

On the other hand, the fast-developing field of laser-nucleus interactions faces important scientific challenges at the intersection of strong field quantum electrodynamics (QED) and nuclear physics. In previous studies, researchers focused more on the nucleus \(^{229}\text{Th}\) in laser-nuclear interactions, especially in experimental studies. This is due to the ultra-narrow nuclear transition energy (8.36 eV) of \(^{229}\text{Th}\), making it a prime candidate for next-generation nuclear clocks \cite{PhysRevResearch.7.L022036,PhysRevLett.128.043001,PhysRevLett.134.113801,PhysRevResearch.7.013052}. However, \(^{229}\text{Th}\) is extremely rare in nature so that it is impossible to be used widely in laboratories. Another atomic nucleus with an isomer state below 100 eV is \(^{235}\text{U}\), whose isomer state has a long spontaneous radiation lifetime, making it a stable system for studying the mechanism of laser-induced nuclear excitation. \cite{BROWNE2014205,PhysRevC.97.054310}. The recent studies by Qi et al. have shown that laser-matter interactions can generate electrons with energies on the 100 eV magnitude, providing strong support for investigating \(^{235}\text{U}\) isomeric state in experiments \cite{PhysRevC.110.L051601}. It is worth noting that the current laser intensity can easily accelerate electrons to keV energy, which far exceeds the threshold required for isotope excitation, enabling full investigation of the excitation efficiency of \(^{235}\text{U}\)--particularly within the non-relativistic regime, which is crucial for the effective control of nuclear excitation as it avoids relativistic effects.  

Previous studies on nuclear excitation via inelastic electron scattering in \(^{235}\text{U}\) have predominantly concentrated on low-energy ranges and specific excitation channels, such as the direct excitation from the ground state (GS) to the isomeric state (IS) \cite{PhysRevC.106.064604}. However, it is essential to recognize that the NEIES encompasses various competing channels that depend on electron energies. By adjusting the energies of the electrons, it is found that one can optimize the excitation channels to enhance the excitation efficiency through a simple two-step process: first, exciting \(^{235}\text{U}\) to higher-energy states, and then allowing it to decay to the IS.  This method is based on the energy-dependent characteristics of the excitation cross-sections, where specific electron energy ranges can be targeted to enhance the population of the higher excited states, thereby increasing the production efficiency of isomers.

Based on the above discussion, this study combines numerical simulations with theoretical calculations to investigate efficient multi-channel excitation of $^{235}\text{U}$ and proposes an efficient indirect excitation scheme for obtaining $^{235}\text{U}$ isomers. This paper is organized as follows. Sec. \ref{section 2} elaborates on nuclear excitation via inelastic electron scattering, encompassing the theoretical framework for calculating inelastic electron scattering cross sections and excitation rates as well as the analysis of excitation cross-section properties and cascading nuclear excitation efficiency. Sec. \ref{section 3} focuses on PIC simulations of laser-cluster interactions, including the simulation methods involving laser parameters and cluster models and the discussion of simulation results regarding laser polarization effects on electron distribution and \(^{235}\text{U}\) nuclear excitation efficiency. Finally, Sec. \ref{section 4} provides a brief summary.

\section{Nuclear Excitation via Inelastic Electron Scattering }
\label{section 2}
\subsection{Theoretical Framework } 

In the context of investigating nuclear excitation of $^{235}\text{U}$ via inelastic electron scattering, the electron potentials can be derived using the Dirac-Hartree-Fock-Slater (DHFS) method \cite{SALVAT2019165}, by considering the nuclear, electronic, and exchange contributions. It is given by   
\begin{equation}
\label{eq_vdhfs}
V_{\mathrm{DHFS}}(r) = V_{\mathrm{n}}(r) + V_{\mathrm{e}}(r) + V_{\mathrm{ex}}(r).
\end{equation}  
The nuclear potential \(V_{\mathrm{n}}(r)\) stems from a Fermi-distributed nuclear charge density \cite{PhysRev.101.1131}, which can be written as  
\begin{equation}
\label{eq_vn}
V_{\mathrm{n}}(r) = -\frac{1}{r} \int_0^r \rho_{\mathrm{n}}(r) 4 \pi r^2 dr - \int_r^\infty \rho_{\mathrm{n}}(r) 4 \pi r dr.
\end{equation} 
Here, $\rho_{\mathrm{n}}(r)$ refers to the nuclear charge density, which can be expressed as
\begin{equation}
\label{eq_rhon}
\rho_{\mathrm{n}}(r) = \frac{\rho_0}{\exp\left[(r-R_{\mathrm{n}})/z\right] + 1}, 
\end{equation}  
with \(A\) the mass number, $\quad R_{\mathrm{n}} = 1.07 A^{1/3}\ \text{fm}$ the nuclear radius, and \(\rho_0\) a normalization constant.

The electronic potential \(V_{\mathrm{e}}(r)\) depends on the electron density \(\rho_e(r)\) from occupied states. For general cases, this density is given by 
\begin{equation}
\label{eq_rhoe1}
\rho_e(\mathbf{r}) = \sum \phi_{n l j m}^{\dagger}(\mathbf{r}) \phi_{n l j m}(\mathbf{r}),
\end{equation}  
and can be simplified for closed shells to
\begin{equation}
\label{eq_rhoe2}
\rho_e(r) = \frac{1}{4 \pi} \sum_{n l j} q \left[ g_{n l j}^2(r) + f_{n l j}^2(r) \right],
\end{equation}  
where \(q\) is the occupation number, and \(g_{n l j}(r)\), \(f_{n l j}(r)\) are radial components of Dirac spinors. Using this density, the electronic potential can be written as 
\begin{equation}
\label{eq_ve}
V_{\mathrm{e}}(r) = \frac{1}{r} \int_0^r \rho_e(r) 4 \pi r^2 dr + \int_r^\infty \rho_e(r) 4 \pi r dr.
\end{equation}

The exchange potential \(V_{\mathrm{ex}}(r)\) uses the Thomas-Fermi approximation \cite{PhysRev.81.385}, which can be written as  
\begin{equation}
\label{eq_vex}
V_{\mathrm{ex}}(r) = 
\begin{cases} 
C_{\mathrm{ex}} V_{\mathrm{ex}}^{(\mathrm{TF})}(r), & r < r_{\text{Latter}} \\
(N+1-Z)\frac{1}{r}-V_{\mathrm{n}}(r)-V_{\mathrm{e}}(r), & r \geq r_{\text{Latter}}
\end{cases},
\end{equation}  
with \(C_{\mathrm{ex}} = 1.5\) \cite{PhysRev.99.510}, a value consistent with the variational treatment of the non-relativistic Hartree-Fock exchange potential under the Thomas-Fermi approximation, where $Z$ nd $N$ are the atomic number and electron count, respectively. $V_{\mathrm{ex}}^{(\mathrm{TF})}(r)$ is the exchange potential
for a free-electron gas. For inelastic scattering, electronic states are modeled as Dirac distorted-wave continuum states \cite{PhysRevLett.124.242501}, which can be written as  
\begin{equation}
\label{eq_phi}
\begin{aligned}
|\phi^{\pm}\rangle &= \frac{4 \pi}{p} \sqrt{\frac{E+c^2}{2 E}} \\
& \times \sum_{j l m} \left[ \Omega_{j l m}^*(\mathbf{v}) \chi_v \right] e^{\pm i \delta_{j l}} \binom{g_{j l}(r) \Omega_{j l m}(\hat{r})}{-i f_{j l'}(r) \Omega_{j l m}(\hat{r})}.
\end{aligned}
\end{equation}  
Here \(\mathbf{v}\) and \(\chi_v\) are velocity/spin vectors, \(\delta_{j l}\) is the phase shift, and \(\Omega_{j l m}(\hat{r})\) are spherical spinors.

The electron-nucleus interaction Hamiltonian includes current-vector potential coupling and Coulomb terms:  
\begin{equation}
\label{eq_h}
\begin{aligned}
H_{\mathrm{int}} &= -\frac{1}{c} \int \left[\boldsymbol{j}_n(\boldsymbol{r}) + \boldsymbol{j}_e(\boldsymbol{r})\right] \cdot \mathbf{A}(\boldsymbol{r}) d \tau \\
& + \int \frac{\rho_n(\boldsymbol{r}) \rho_e(\boldsymbol{r}')}{\left|\boldsymbol{r}-\boldsymbol{r}'\right|} d \tau d \tau',
\end{aligned}
\end{equation}  
with \(\boldsymbol{j}(\boldsymbol{r})\)/ \(\rho(\boldsymbol{r})\) the current/charge densities, and \(\mathbf{A}(\boldsymbol{r})\) the vector potential.

The matrix element \(\langle f | H_{\mathrm{int}} | i \rangle\) is expanded via electric \((E\lambda)\)/magnetic \((M\lambda)\) multipole transitions \cite{RevModPhys.30.353}, which can be given by
\begin{equation} 
\label{eq_matrix} 
\begin{aligned} 
\langle f | H_{\mathrm{int}} | i \rangle &= \sum_{\lambda \mu} \frac{4 \pi}{2 \lambda+1}(-1)^\mu \\
& \times \left\{ \left\langle\phi_f \left| \mathcal{N}(E\lambda, \mu) \right| \phi_i \right\rangle \left\langle I_f M_f \left| \mathcal{M}(E\lambda,-\mu) \right| I_i M_i \right\rangle \right. \\
& \left.-\left\langle\phi_f \left| \mathcal{N}(M\lambda, \mu) \right| \phi_i \right\rangle \left\langle I_f M_f \left| \mathcal{M}(M\lambda,-\mu) \right| I_i M_i \right\rangle \right\}.
\end{aligned}
\end{equation}  
Here \(\mathcal{M}/\mathcal{N}\) are nuclear/electronic multipole operators, and \(\left|I M\right\rangle/\left|\phi\right\rangle\) are nuclear/electronic states.

The differential cross section for electron-nucleus energy transfer, derived via Fermi's golden rule, is given by
\begin{equation}
\label{eq_diff1}
\frac{d \sigma}{d \Omega} = \frac{2 \pi}{v_i} \rho(E_f) \left|\langle f | H_{\mathrm{int}} | i \rangle\right|^2,
\end{equation}  
with \(v_i\) the incident electron speed and \(\rho(E_f)\) the final-state density. Incorporating reduced transition probabilities \(B(E(M)\lambda; I_i \to I_f)\),  Eq.(\ref{eq_diff1}) can be written as  
\begin{equation}
\label{eq_diff2}
\begin{aligned}
\frac{d \sigma}{d \Omega} &= \frac{4E_i E_f}{c^4} \frac{p_f}{p_i} \sum_{\lambda E(M) \mu} \left\{ \frac{B(E(M)\lambda; I_i \to I_f)}{(2 \lambda + 1)^3} \right. \\
& \left. \times \frac{1}{2} \sum_{v_i v_f} \left| \left\langle \phi_f^{-} \left| \mathcal{N}(E(M)\lambda, \mu) \right| \phi_i^{+} \right\rangle \right|^2 \right\}.
\end{aligned}
\end{equation}  
In the above expression, \(E\) and \(p\) are relativistic electron energy and momentum. $\lambda$ and $\mu$ are the angular momentum and magnetic quantum
number, respectively.

The total excitation rate for NEIES in a system with \(N_{\text{tar}}\) nuclei is given by the product of the per-nucleus excitation rate \(\lambda_{\mathrm{NEIES}}\) and the total number of target nuclei. It is given by 
\begin{equation}
\label{eq_total_rate}
N_{\text{NEIES}} = \lambda_{\text{NEIES}} \times N_{\text{tar}},
\end{equation}  
where \(\lambda_{\mathrm{NEIES}}\) is obtained by integrating the sum of electric (\(\sigma_{\mathrm{E}\lambda}\)) and magnetic (\(\sigma_{\mathrm{M}\lambda}\)) cross sections over the electron energy spectrum, weighted by the electron flux \(\Phi_e(E)\). Thus, it can be written as
\begin{equation}
\label{eq_per_rate}
\lambda_{\mathrm{NEIES}} = \int dE \left[ \sigma_{\mathrm{E}\lambda}(E) + \sigma_{\mathrm{M}\lambda}(E) \right] \Phi_e(E).
\end{equation}  
The total NEIES cross section combining these components \cite{PhysRevLett.124.242501} can be given by  
\begin{equation}
\label{eq_sigma}
\begin{aligned}
\sigma_{\mathrm{E(M)\lambda}} &= \frac{8\pi^2}{c^4} \frac{E_i + c^2}{p_i^3} \frac{E_f + c^2}{p_f} \\
& \times \sum_{l_i, j_i, l_f, j_f} \frac{\kappa^{2\lambda+2}}{(2\lambda-1)!!^2} B(E(M)\lambda, I_i \to I_f) \\
& \times \frac{(2l_i+1)(2l_f+1)(2j_i+1)(2j_f+1)}{(2\lambda+1)^2} \\
& \times \left( \begin{array}{ccc} l_f & l_i & \lambda \\ 0 & 0 & 0 \end{array} \right)^2 \left\{ \begin{array}{ccc} l_i & \lambda & l_f \\ j_f & 1/2 & j_i \end{array} \right\}^2 \left| T_{f i}^{E(M)\lambda} \right|^2,
\end{aligned}
\end{equation}  
with radial matrix elements for electric and magnetic transitions:
\begin{equation}
\label{eq_te}
T_{f i}^{E\lambda} = \int_0^\infty h_\lambda^{(1)}(\kappa r) \left[ g_i(r)g_f(r) + f_i(r)f_f(r) \right] r^2 dr
\end{equation}  
and  
\begin{equation}
\label{eq_tm}
\begin{aligned}
T_{f i}^{M\lambda} &= \frac{\eta_i + \eta_f}{\lambda} \\
& \times \int_0^\infty h_\lambda^{(1)}(\kappa r) \left[ g_i(r)f_f(r) + g_f(r)f_i(r) \right] r^2 dr.
\end{aligned}
\end{equation}  
Here \(\eta = (l-j)(2j+1)\), \(h_\lambda^{(1)}\) are spherical Hankel functions of the first kind, and \(\kappa\) corresponds to the excitation energy divided by the speed of light.

To quantitatively characterize the excitation of \(^{235}\text{U}\) nuclei from the GS to excited states through inelastic electron scattering processes, our derivation is based on atomic units (a.u.). In this unit system, fundamental constants such as \(\hbar\), \(m_e\) (electron mass), and \(e\) (elementary charge) are set to 1, with lengths in Bohr radii and energies in Hartrees. The speed of light is \(c = 1/\alpha \approx 137.036\) ( \(\alpha\) is the fine-structure constant). Initial and final states are labeled by subscripts \(i\) and \(f\); \(n\) and \(e\) distinguish nuclear and electronic quantities. Electron quantum numbers--principal \((n)\), orbital angular momentum \((l)\), total angular momentum \((j)\), and magnetic \((m)\)--are defined to describe interactions and compute cross sections systematically.

\subsection{Excitation cross-sections of \(^{235}\text{U}\) and cascading nuclear excitation}

\begin{figure}[H]
\centering
\includegraphics[width=8cm]{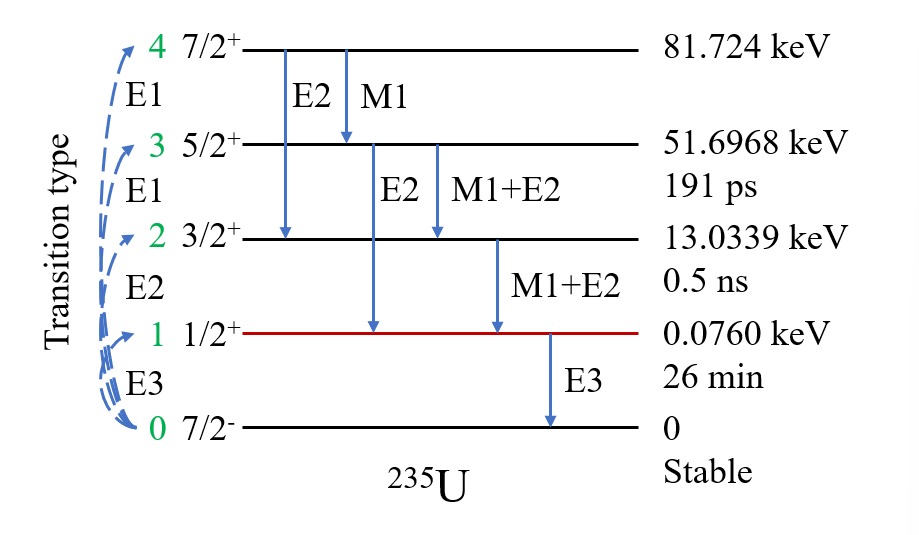}
\caption{Energy level diagram of \(^{235}\text{U}\) with different nuclear excited states and decay processes to the isomeric state.} 
\label{fig 1}
\end{figure}

To clearly illustrate the process of indirect excitation, the different excitation and de-excitation channels of \(^{235}\text{U}\) are shown in Fig. \ref{fig 1}. As illustrated in this figure, the \(^{235}\text{U}\) nucleus has distinct states with specific spin-parities. Following the selection rules for multipole radiation and parity conservation, transitions from the ground state (state \(0\), \(7/2^-\), \(0\)) to other excited states occur via corresponding multipole modes. Specifically, the transition from the ground state to the first excited state (state \(1\), \(1/2^+\), \(76\ \text{eV}\)) proceeds through the electric octupole ($E3$) mode. The transition to the second excited state (state \(2\), \(3/2^+\), \(13.0339\ \text{keV}\)) occurs via the electric quadrupole ($E2$) mode. The transitions to the third (state \(3\), \(5/2^+\), \(51.6968\ \text{keV}\)) and fourth (state \(4\), \(7/2^+\), \(81.724\ \text{keV}\)) excited states both proceed through electric dipole ($E1$) modes. In terms of de-excitation from the excited state to the isomeric state, the \(^{235}\text{U}\) nuclear level structure reveals distinct characteristics from other nuclei, i.e., the isomeric state has a long half-life of 26 minutes. This stability arises from the low probability of high-multipolarity $E$3 transitions, making the IS ideal for long-term experimental observation. The second excited state decays to the IS via magnetic dipole ($M1$) or $E2$ transitions with a half-life of 0.5 ns, while the third excited state exhibits a 191 ps half-life and decays through $M1$+$E2$ transitions. These short-lived states serve as intermediaries in indirect excitation, directing the population to the isomer state (IS) through rapid decay. The lifetime of 26 minutes is significantly longer than the lifetimes of higher excited states, which range from nanoseconds to picoseconds. This stability allows for the efficient accumulation of IS nuclei. By simultaneously analyzing both direct and indirect excitation channels, this study shows that before electrons reach relativistic speeds (approximately 100 keV), indirect excitation through higher excited states can enhance excitation efficiency by eleven orders of magnitude compared to direct excitation alone. This conclusion will be further discussed in the subsequent discussion.

\begin{figure*}[hbt]
\centering
\includegraphics[width=16cm]{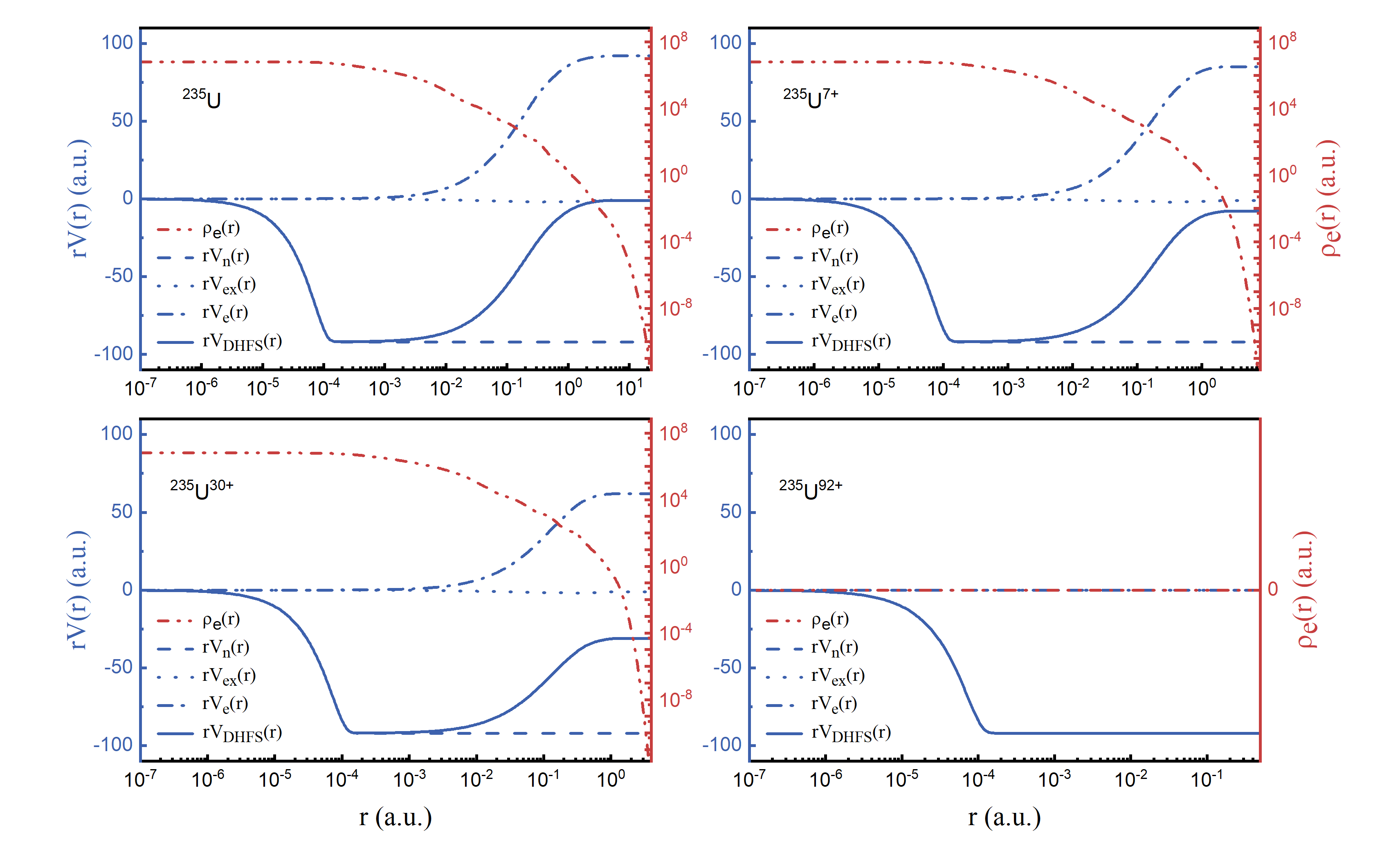}
\caption{DHFS self-consistent potential $rV_{\mathrm{DHFS}}(r)$ for the $^{235}\text{U}$ atom and selected ions \(^{235}\)U\(^{7+}\), \(^{235}\)U\(^{30+}\) and \(^{235}\)U\(^{92+}\). The DHFS radial electron density \(\rho_e(r)\) is represented by the red dashed line. The nuclear potential (\(rV_{\mathrm{n}}\)), electronic potential (\(rV_{\mathrm{e}}\)), exchange potential (\(rV_{\mathrm{ex}}\)), and total potential (\(rV_{\mathrm{DHFS}}\)) are denoted by different blue dashed lines, respectively. }
\label{fig 2}
\end{figure*}

This work focuses on the \(^{235}\text{U}\) atom and its ions in various charge states, using the DHFS method to calculate potential energy distributions and electron densities. We calculated the potential energy curves and radial electron density profiles for a neutral $^{235}\text{U}$ atom, as well as for \(^{235}\)U\(^{7+}\), \(^{235}\)U\(^{30+}\) and \(^{235}\)U\(^{92+}\) ions in Fig. \ref{fig 2}. These specific charge states are chosen to understand how charge states affect radial wave functions, which are crucial for determining excitation cross-sections in PIC simulations.

As shown in Fig. \ref{fig 2}, the nuclear potential curves of the atom and all ionic states are nearly identical, as it is solely determined by the proton charge and nuclear size.   For different ionic states, the exchange potential's energy curve is only slightly affected by the degree of ionization, whereas the electronic potential's energy curve decreases significantly as ionization increases. For the highly ionized ion U$^{92+}$, the electronic potential becomes negligible due to the absence of bound electrons. This behavior is closely related to the electron density distribution shown by the red curve in Fig. \ref{fig 2}. The electron density reaches a maximum near the nucleus and gradually decreases, indicating strong electron shielding. As ionization increases, electron density localizes closer to the nucleus, ultimately vanishing entirely for U$^{92+}$. 

Thus, the total potential evolves distinctively in neutral and ionization states. The extended electron cloud leads to a gradual decay of the total potential at larger radii. In contrast, highly ionized states exhibit sharp potential variations near the nucleus, dominated by the unscreened nuclear Coulomb field. These structural variations in the potential directly influence the radial wave functions, emphasizing the necessity of considering different ionic states when calculating the NEIES excitation channels of \(^{235}\text{U}\) in the following PIC simulations.

To further explore the excitation mechanisms of \(^{235}\text{U}\), we examine the nuclear states and their properties. The reduced transition probability from the IS to the GS is \(B(\text{$E3$}) = 0.036\ \text{W.u.}\) \cite{PhysRevLett.121.253002}, while for other transitions, the \(B\) values are all set to \(1\  \text{W.u.}\), consistent with the value by Feng et al. \cite{pnas2413221121}. We calculate the excitation cross sections of \(^{235}\text{U}\) over appropriate ranges of incident electron energies for these transitions and focus on the nuclear excited states as shown in Fig. \ref{fig 1}. On one hand, the complex decay channels of excitations beyond these states contribute negligibly to the accumulation of nuclei in the IS (\(1/2^+\)). On the other hand, electron energies exceeding those corresponding to these transitions will reach a level where relativistic effects need to be considered. This will lead to significant changes in the electron-nucleus interaction dynamics, such as altering the effective potential experienced by electrons, modifying the multipole transition probabilities in a complex way, and complicating the calculation of excitation cross sections due to the need to incorporate relativistic quantum mechanics formulations. The calculations for these channels are performed for the neutral \(^{235}\text{U}\) atom and will be extended to several ionized states.  

\begin{figure*}[hbt]
\includegraphics[width=18cm]{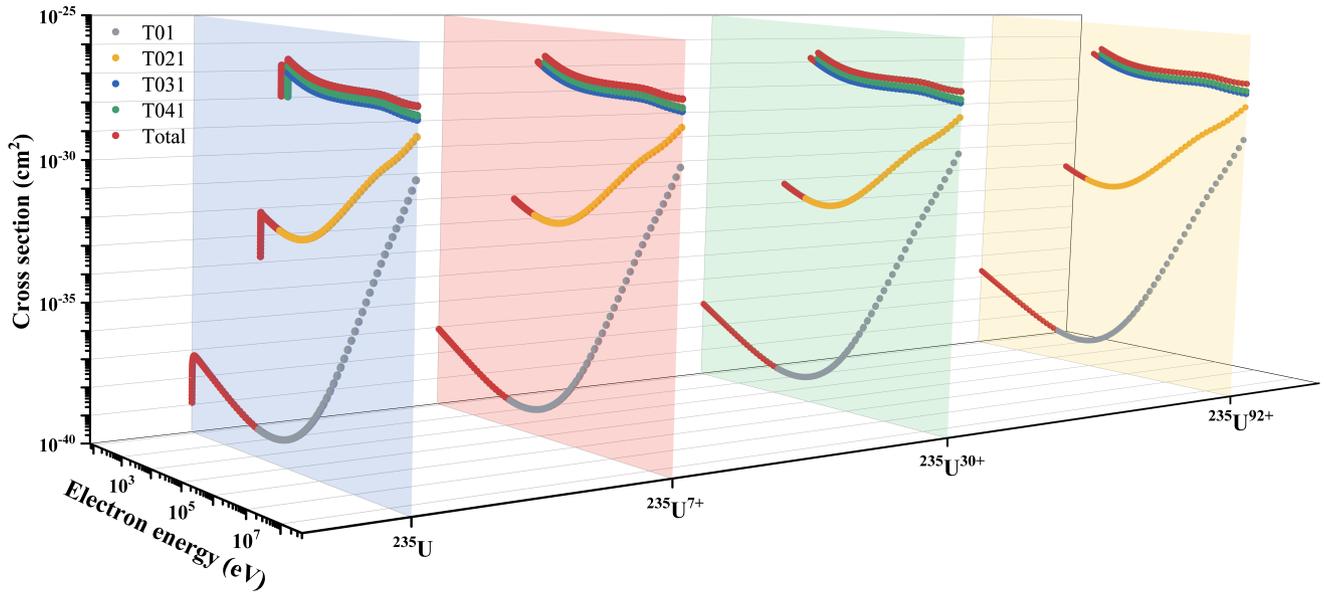}
\caption{Reaction cross sections of different excitation channels for \(^{235}\text{U}\) atom, \(^{235}\)U\(^{7+}\), \(^{235}\)U\(^{30+}\) and \(^{235}\)U\(^{92+}\) ions versus the electron energy.}
\label{fig 3}
\end{figure*}

\begin{figure}[H]
	\centering
	\includegraphics[width=8cm]{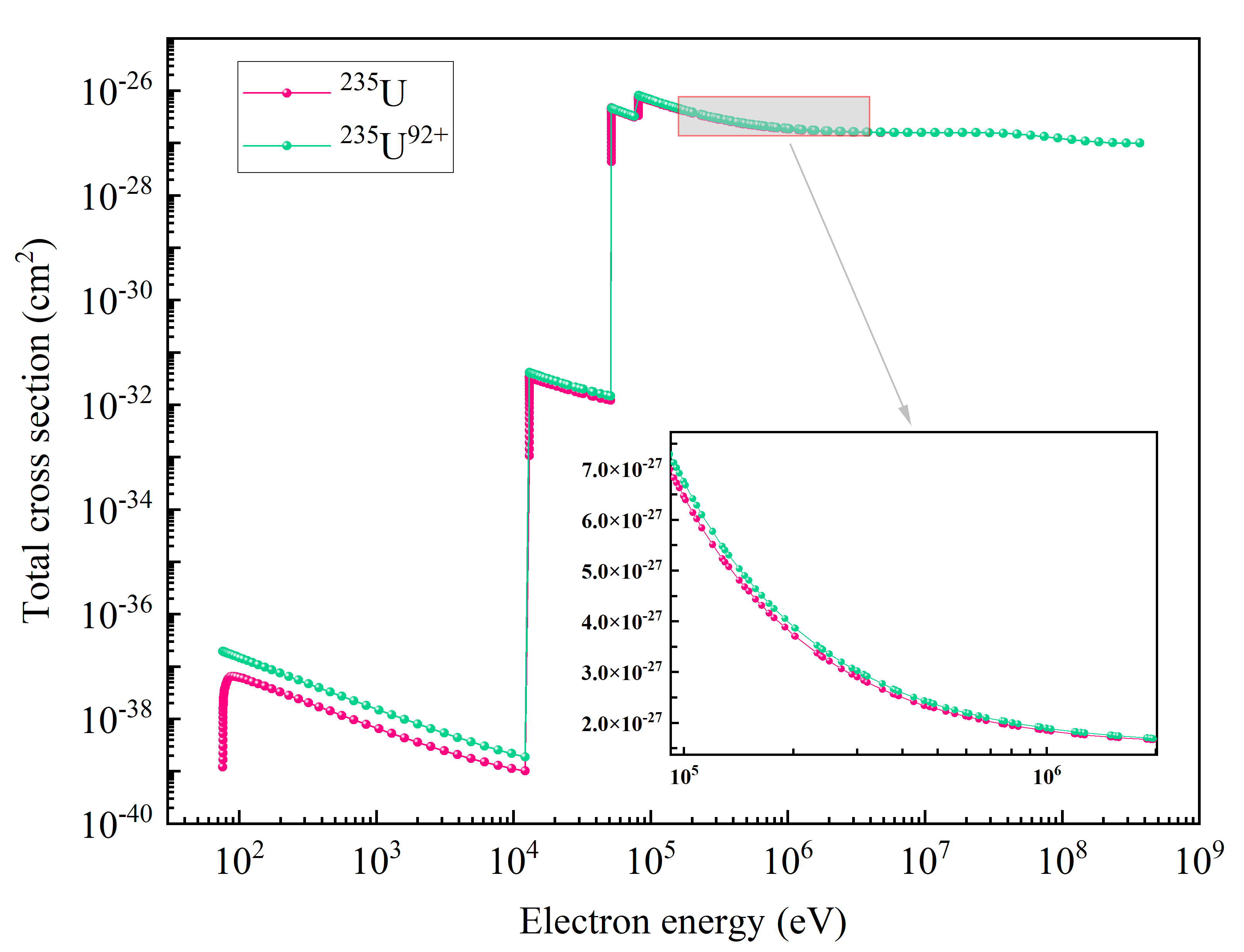}
	\caption{Total cross sections of different excitation channels for \(^{235}\text{U}\) atom and \(^{235}\)U\(^{92+}\) ions versus electron energy.} 
	\label{fig 444}
\end{figure}

As shown in the Fig. \ref{fig 3}, the reaction cross-sections for various excitation channels (T01, T021, T031, and T041, which correspond to transitions from the ground state to the 1st through 4th excited states, respectively) and the total cross-section for neutral \(^{235}\text{U}\) atoms demonstrate significant energy-dependent characteristics. In the low-energy region (approximately \(10^2-10^4\ \text{eV}\)), the cross section of the T01 channel rapidly reaches its peak near the threshold energy (approximately \(10^2\ \text{eV}\)), and then begins to decline as the electron energy increases. Upon entering the medium-energy region (approximately \(10^4-10^6\ \text{eV}\)), the cross-section of the T01 channel increases significantly with rising energy. The T021-T041 channels begin to activate, and the total reaction cross-section reaches a peak at an electron energy of approximately 81 keV. For the ionic states \(^{235}\)U\(^{7+}\), \(^{235}\)U\(^{30+}\) and \(^{235}\)U\(^{92+}\), calculations show that their reaction cross-sections have the same trend as that of neutral \(^{235}\text{U}\) atoms, but there are subtle differences in values. To more clearly illustrate these subtle differences, we have plotted the total reaction cross-sections corresponding to the \(^{235}\text{U}\) atom and the fully ionized \(^{235}\)U\(^{92+}\) ion in Fig. \ref{fig 444}. As can be seen from Fig. \ref{fig 444}, there are slight differences between the two in the low-energy region (\(10^2\text{-}10^4\ \text{eV}\)), with the cross-section peak of \(^{235}\)U\(^{92+}\) being slightly higher than that of neutral \(^{235}\text{U}\). This may be related to the change in the interaction efficiency between low-energy electrons and the nucleus due to the absence of electron shielding after full ionization. In the medium-to-high energy region (above \(10^4\ \text{eV}\)), as the electron energy increases, the influence of the charge state on the scattering process weakens, and the cross-section curves of the two gradually overlap. This is because the electron shielding effect can be neglected relative to the effect of high-energy electrons at this point.

Combined with the previous analysis of nuclear state transitions and cross-sections, the cascading nuclear excitation mechanism by high-energy electrons can significantly improve the IS's production efficiency due to the advantage of cross-section accumulation in indirect excitation. Although the direct excitation cross section of the IS  (T01 channel) has a peak in the low-energy region, it rapidly decays with increasing electron energy. If high-energy electrons are used first to excite these high-energy states, and then they cascade back to the IS  through transitions (such as high-energy state 3/4 $\to$ excited state 2 $\to$ IS), the cross-section contributions of multiple channels can be accumulated. For example, the cross-sections of the T031 and T021 channels in the \(10^6-10^7\ \text{eV}\) region are much higher than the T01 channel, and the cascading process can convert the cross-sections of these high-energy channels into the effective accumulation of the IS.  

Since a large number of ionic states $^{235}\text{U}$ are generated in PIC simulations, the coexistence of multiple ionic states can further improve the production efficiency of the IS  through `hybrid channel excitation'. Low-energy electrons can directly excite nuclei to the IS, while high-energy electrons can excite high-energy states and return through cascading to replenish the IS, making the accumulation efficiency of the IS much higher than that of single direct excitation.

In conclusion, the efficiency bottleneck of direct excitation can be overcome by regulating the electron energy, utilizing a cascading transition channel, and hybrid channel excitation in a laser-driven cluster environment. This approach allows for the effective accumulation of IS nuclei. These insights provide a theoretical foundation for optimizing the NEIES excitation scheme and increasing the proportion of IS in future experiments.

\section{PIC Simulations of Laser-Cluster Interactions}
\label{section 3}
\subsection{Methods}

In the present work, we perform the two-dimensional three-velocity (2D3V) PIC simulations with the open-source code EPOCH \cite{Arber_2015} to model the laser-cluster interaction dynamics. The size of the simulation box is $1\ \mu\text{m} \times 1\ \mu\text{m}$ with a spatial resolution of $1\ \text{nm} \times 1\ \text{nm}$ in x and y direction, respectively, ensuring sufficient resolution to capture the nanoscale cluster dynamics. This configuration enables self-consistent modeling of electromagnetic field propagation, charged particle motion, and ionization processes within the $^{235}\text{U}$ cluster.

The linearly polarized laser pulse utilized in the PIC simulation is characterized by a Gaussian temporal profile, featuring a full width at half maximum (FWHM) of 20 fs. Key simulation parameters include peak intensities of  \(1 \times 10^{17}\ \text{W/cm}^2\) and \(1 \times 10^{18}\ \text{W/cm}^2\), and a wavelength of 800 nm. These parameters are selected to investigate how nuclear excitation mechanisms depend on laser parameters under consideration. Furthermore, both laser-field ionization and collisional ionization are incorporated in all simulations. While incorporating collisional ionization significantly increases computation time, it ultimately provides more reliable results due to the complex interactions among the lasers, electrons, and nuclei.


In the PIC simulations, the cluster density of \(^{235}\text{U}\) is a critical factor, as it has a direct impact on the dynamics of electron-nucleus interactions. In the present work, a \(^{235}\text{U}\) cluster with a radius of \(20\ \text{nm}\) and a number density of \(n_0 = 2.35 \times 10^{22}\ \text{cm}^{-3}\) is considered. This configuration corresponds to a cluster containing \(N_{\text{tar}}\) atoms, which can be calculated from the volume of the sphere and its density: $N_{\text{target}}= \frac{4}{3}\pi R^3 \cdot n_0 = \frac{4}{3}\pi (20\ \text{nm})^3 \cdot 2.35 \times 10^{22}\ \text{cm}^{-3}\approx 7.87 \times 10^5.$
This size ensures that electrons remain sufficiently confined within the cluster during laser-cluster interactions, preventing excessive diffusion that would reduce the probability of electron-nucleus collisions.

\subsection{Results and Discussion}

Previous PIC simulations of laser-cluster interactions for \(^{235}\text{U}\) have primarily focused on electron energies below 100 eV, where direct excitation to the IS (76.7 eV) dominates \cite{PhysRevC.110.L051601}. However, our calculations show that exciting the \(^{235}\text{U}\) atom to the fourth excited state requires a minimum electron energy of 81.724 keV, which necessitates adjusting the laser parameters in PIC simulations. The main challenge is to generate high-density electron beams with an energy of 100 keV under laser intensities below \(1 \times 10^{18}~\text{W/cm}^2\) (the threshold for relativistic electron dynamics). This avoids complex relativistic effects while achieving efficient nuclear excitation.

\begin{figure}[H]
	\centering
	\includegraphics[width=8cm]{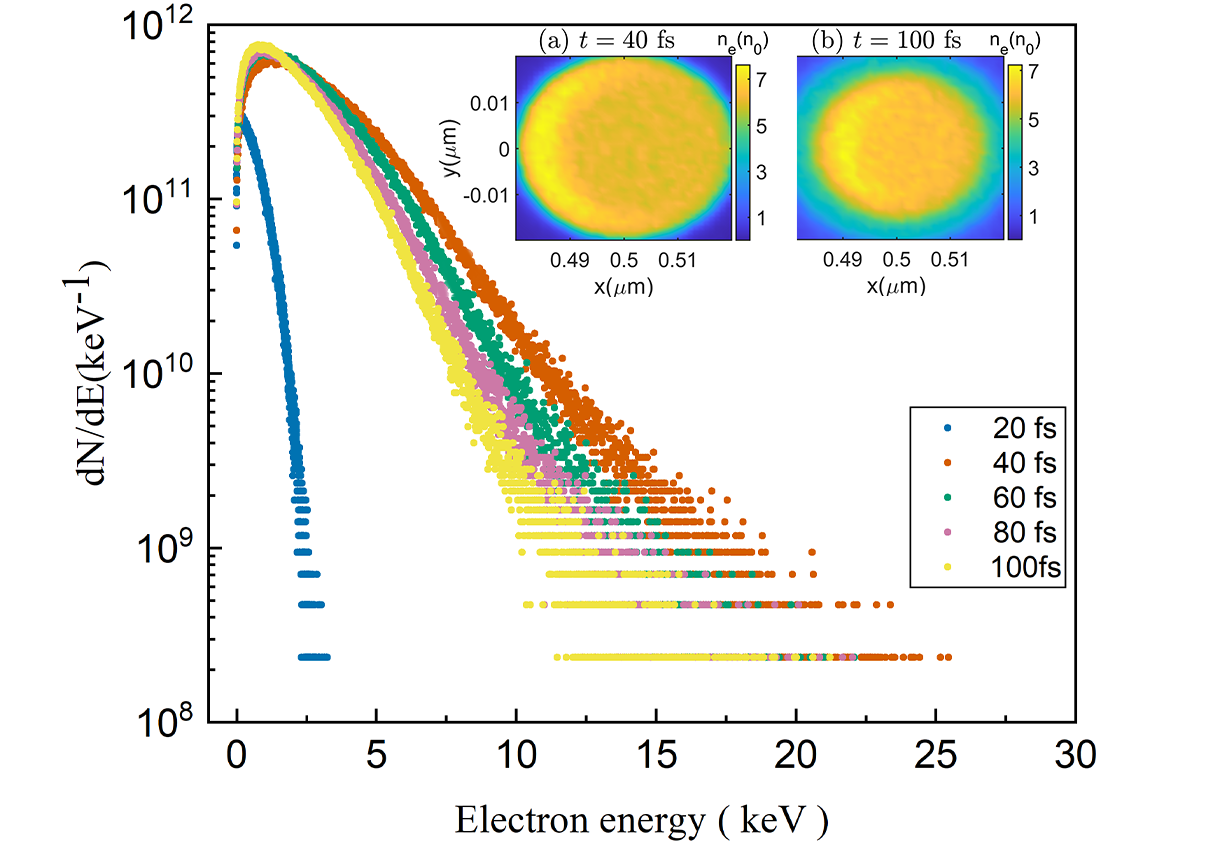}
	\caption{Energy spectra evolution of electrons. Insets (a) and (b) display the electron density distribution of $^{235}\text{U}$ clusters irradiated by a \(10^{17}\ \text{W/cm}^2\) s-polarized laser pulse at $t = 40\ \text{fs}$ and at $t = 100\ \text{fs}$, respectively. }
\label{fig 5}
\end{figure}

Here, we first simulate the electron density and energy distributions of clusters ionized by a \(10^{17}\ \text{W/cm}^2\) s-polarized linear laser pulse. As depicted in the Fig. \ref{fig 5} (a), at \(t = 40\ \text{fs}\), a partial region of the cluster exhibits an electron density exceeding \(7n_0\) (\(n_0\) is the atomic number density of $^{235}\text{U}$), indicating a significant ionization enhancement in the core area due to the laser field ionization. By \(t = 100\ \text{fs}\) (Fig. \ref{fig 5} (b)), the electrons begin to exhibit diffusion phenomena. Those electrons initially concentrated in the high-density regions begin to diffuse outward radially, leading to a gradual decrease in the electron density gradient between the center and the edge. This behavior aligns with the mechanisms of diffusion that occur when electrons have lower energy during laser-cluster interactions \cite{PhysRevC.110.L051601}. The key difference is that electrons with an energy level of 10 keV diffuse significantly faster than those with 100 eV, directly affecting the spatial distribution of electrons in clusters and the efficiency of their interactions. Notably, even as the cluster expands, most electrons remain around the core, allowing the high-energy electrons to fully excite $^{235}\text{U}$ during the cluster expansion.  

The electron energy distribution, as shown in Fig. \ref{fig 5}, shows a peak at \(40\ \text{fs}\). However, the electron energy at this moment remains below \(100\ \text{keV}\), which is necessary for exciting the fourth-excited state of \(^{235}\text{U}\). In fact, most of the electrons are concentrated below \(20\ \text{keV}\). This indicates that while the laser intensity is sufficient to induce significant ionization, there needs to be further optimization of the laser parameters to produce a greater number of electrons with energies in the \(100\ \text{keV}\) range. This enhancement is crucial for effectively achieving indirect excitation of \(^{235}\text{U}\) nuclei through higher-energy states.

\begin{figure}[H]
\centering
\includegraphics[width=8cm]{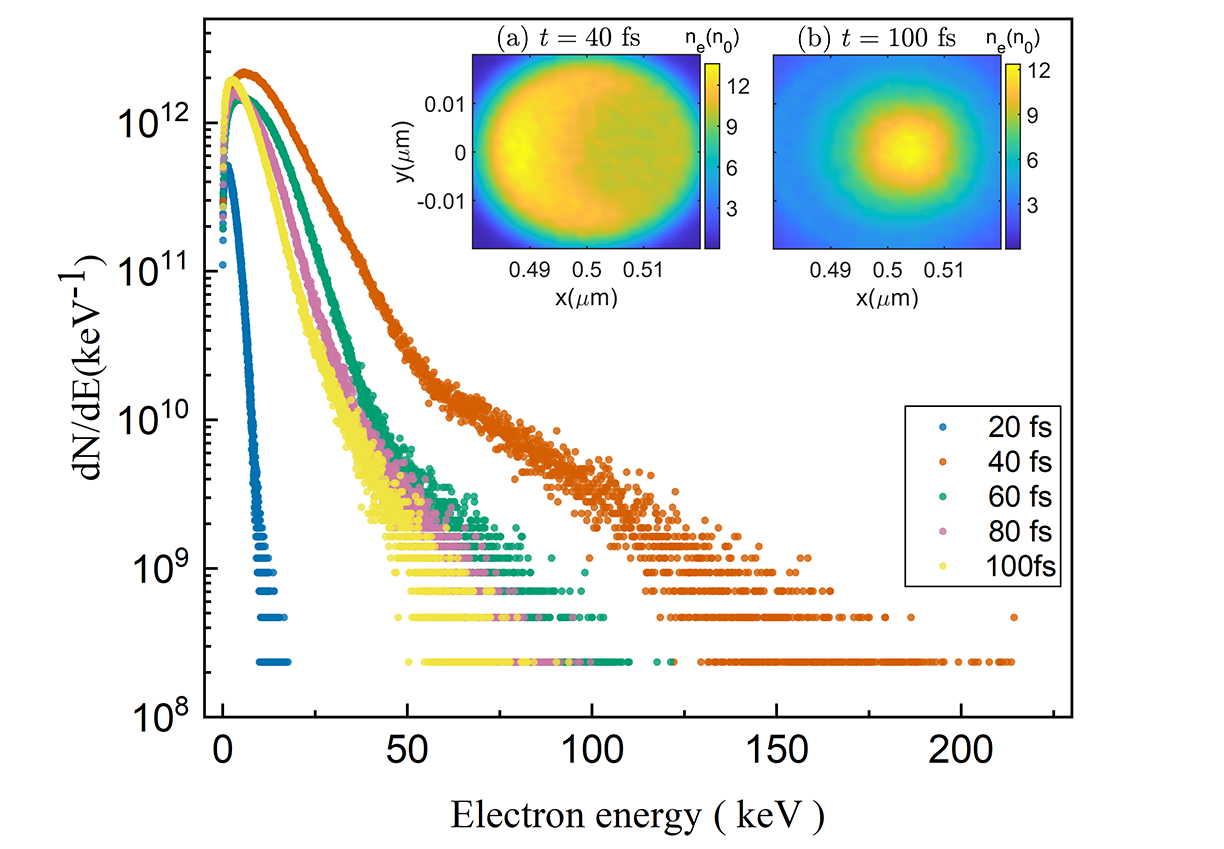}
\caption{ Energy spectra evolution of electrons. Insets (a) and (b) display the electron density distribution of $^{235}\text{U}$ clusters irradiated by a \(10^{18}\ \text{W/cm}^2\) s-polarized laser pulse at $t = 40\ \text{fs}$ and at $t = 100\ \text{fs}$, respectively. }
\label{fig 6}
\end{figure}

To further explore the impact of higher-intensity laser fields on the electron dynamics, we increased the laser intensity to \(10^{18}\ \text{W/cm}^2\) while keeping other parameters constant. At \(t = 40\ \text{fs}\) (Fig. \ref{fig 6} (a)), the electron density distribution shows significant inhomogeneity. This inhomogeneity indicates that for a relativistic laser intensity of \(10^{18}\ \text{W/cm}^2\), the interaction between the laser and the cluster becomes highly complex. Consequently, the straightforward application of models or assumptions valid for lower-intensity fields is no longer applicable. By \(t = 100\ \text{fs}\) (Fig. \ref{fig 6} (b)), the diffusion of electrons becomes extremely fast. Electrons can no longer be well confined within the cluster, which contrasts with the relatively more bounded state at lower laser intensities above. This rapid diffusion implies that the interaction time between the electrons and the \(^{235}\text{U}\) nuclei is significantly reduced, potentially hindering efficient excitation. From the electron energy spectrum (Fig. \ref{fig 6}), we observe that the electron energy reaches a maximum at \(t = 40\ \text{fs}\) and can attain the \(100\ \text{keV}\) magnitude necessary to excite \(^{235}\text{U}\). However, considering the two aforementioned drawbacks--inhomogeneous density distribution and extremely fast diffusion--it is essential to explore experimental strategies for achieving higher electron energies without increasing laser intensity. This approach is vital for the efficient and controllable excitation of \(^{235}\text{U}\) nuclei in the field of NEIES-related research.

\begin{figure}[H]
\centering
\includegraphics[width=8cm]{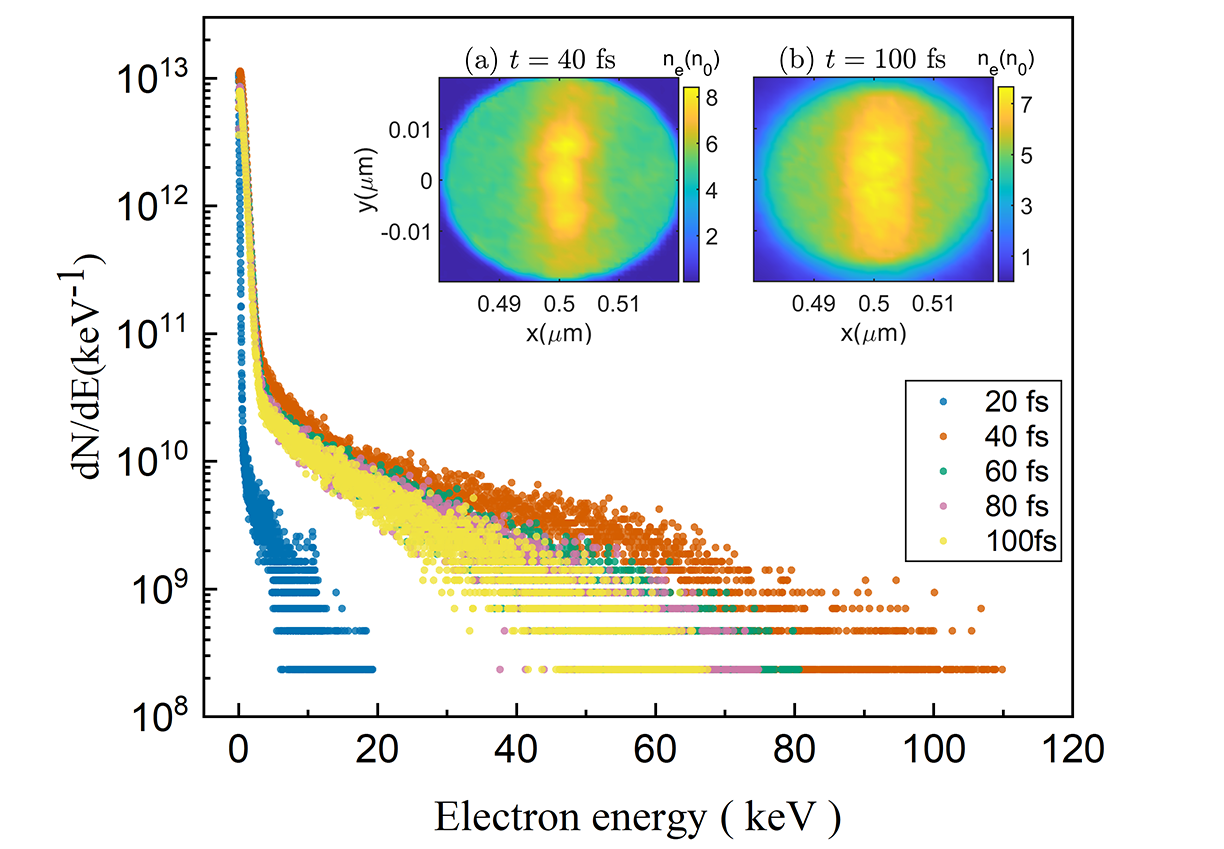}
\caption{Energy spectra evolution of electrons. Insets (a) and (b) display the electron density distribution of $^{235}\text{U}$ clusters irradiated by a \(10^{17}\ \text{W/cm}^2\) p-polarized laser pulse at $t = 40\ \text{fs}$ and at $t = 100\ \text{fs}$, respectively. }
\label{fig 7}
\end{figure}

Considering the polarization dependence of laser-cluster interactions, we used a p-polarized pulse at \(10^{17}\ \text{W/cm}^2\) to investigate the electron-nucleus coupling in \(^{235}\text{U}\). As illustrated in Fig. \ref{fig 7} (a), this configuration showed a pronounced axial concentration of free electrons at \(t = 40\ \text{fs}\), intensifying local density gradients and thereby amplifying the Coulomb interactions with the nuclei. By \(t = 100\ \text{fs}\) (Fig. \ref{fig 7} (b)), most of the electrons remained confined within the cluster, with electron density above \(n_0\) persisting in the core, thus ensuring a longer interaction for the nuclear excitation.

\begin{figure*}[hbt]
\centering
\includegraphics[width=15cm]{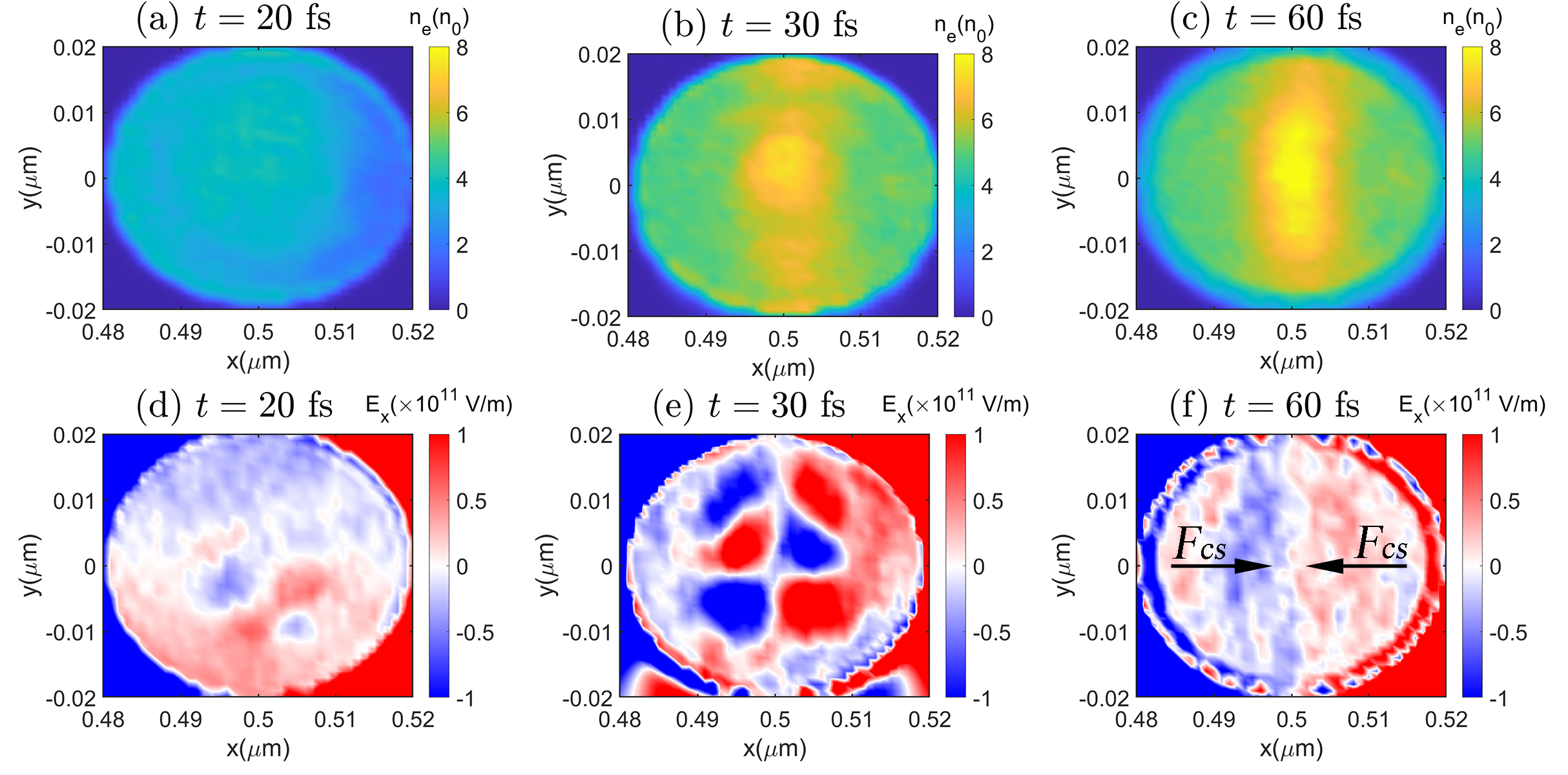}
\caption{Electron density distribution and electric field evolution in \(^{235}\text{U}\) clusters irradiated by a \(10^{17} \, \text{W/cm}^2\) p-polarized laser pulse at \(t=20 \, \text{fs}\), \(t=30 \, \text{fs}\), and \(t=60 \, \text{fs}\), respectively. }
\label{fig 8}
\end{figure*}

The electron energy spectrum  (Fig. \ref{fig 7}) reveals critical dynamic characteristics of \(^{235}\text{U}\) clusters under a \(10^{17}\ \text{W/cm}^2\) p-polarized laser pulse. At early times (\(t \leq 20\ \text{fs}\)), a large number of electrons appear in the 0-10 keV range, and these electrons mainly participate in the direct excitation process of low-energy nuclear states. As time progresses to \(t = 40\ \text{fs}\),  statistics show that the number of high-energy electrons capable of exciting the nucleus to the second excited state is 7 times the amount observed under s-polarized laser conditions. This significant increase directly enhances the probability of nuclear excitation by NEIES occurring via the second state, which undergoes rapid decay to the IS with a 0.5 ns half-life, contributing to efficient IS accumulation.  More importantly, this high-energy component exceeds the threshold for exciting the fourth nuclear state (81.724 keV), laying the foundation for indirect excitation via higher-energy states. By \(t = 100\ \text{fs}\), the low-energy peak decays due to the electron thermalization, while the high-energy tail extends to 90 keV. This sustained high-energy electron flux remains sufficient to drive nuclear transitions, ensuring continuous replenishment of higher excited states that cascade back to the IS.  The above data indicate that this temporal evolution of the electron energy spectrum aligns with the cascading nuclear excitation mechanism, highlighting the role of p-polarized lasers in optimizing electron energy distribution for efficient IS generation.

To further elaborate on the superior performance of p-polarized lasers from the perspective of electron dynamics, the temporal evolution of electron density and associated field effects in the cluster is visualized in Fig. \ref{fig 8}. At the initial time of \( t = 20 \, \text{fs} \), the electron density is observed to be approximately \( 3n_0 \) as shown in Fig. \ref{fig 8} (a). Due to the p-polarized laser being focused precisely at the center of the cluster, the electric field intensity near the cluster core gradually increases, facilitating more efficient ionization of atoms and imparting kinetic energy to the released electrons. This leads to a gradual rise in electron density around the central region of the cluster. By \( t = 30 \, \text{fs} \), the electric field intensity reaches its maximum, driving the electron density near the cluster center to a peak value of \( 8n_0 \) as shown in Fig. \ref{fig 8} (b). Subsequently, while the peak electron density remains stable, the electric field intensity at the cluster center begins to decrease as the laser pulse starts to propagate away from the cluster. By \( t = 60 \, \text{fs} \), the laser pulse has completely exited the cluster, and a charge separation field (\( E_{\text{cs}} \)) is established within the cluster due to the separation of energetic electrons from ions. Under the influence of this charge separation field, electrons experience a confining force along the \( x \)-direction (\( F_{\text{cs}} = -eE_{\text{cs}} \)) as shown in Fig. \ref{fig 8} (f), which effectively confines them in the vicinity of the cluster center. This confinement of electrons in the core region, facilitated by the unique field dynamics induced by p-polarized lasers, ensures prolonged and efficient interactions between electrons and \( ^{235}\text{U} \) nuclei, thereby enhancing the efficiency of isomeric state production.

In conclusion, the energy enhancement induced by changing the laser polarization mode is crucial for the increase in NEIES reaction cross-sections discussed in this study. The 40 fs high-energy peak corresponds with the optimal timing for populating the fourth excited state. This state subsequently undergoes a cascading decay sequence (e.g., \(7/2^+ \to 5/2^+ \to 3/2^+ \to 1/2^+\)), which can effectively accumulate the IS. By concentrating electrons along the axis (\(x = 0.5 \, \mu\text{m}\)) and increasing their high-energy flux, p-polarized lasers provide several advantages -- they maximize the overlap between electrons and the nucleus while aligning energy levels with nuclear transition thresholds. This underscores the importance of laser polarization control as a key factor for tuning the excitation channels of \(^{235}\text{U}\), thereby moving closer to controlled IS accumulation.

\section{Summary}
\label{section 4}

In summary, we investigate the NEIES mechanism in \(^{235}\text{U}\) within laser-irradiated clusters by combining DHFS calculations and 2D3V PIC simulations.  We analyze excitation cross-sections for transitions from the ground state to excited states and validate the efficiency of indirect excitation through the cascading of higher-energy states to the isomeric state. The results show that laser-tuned multi-channel NEIES enhances IS accumulation efficiency by eleven orders of magnitude compared to direct excitation.  PIC simulations show that a p-polarized linear laser with an intensity of \(10^{17}\ \text{W/cm}^2\) can concentrate electrons axially  (at \(x = 0.5 \, \mu\text{m}\)) within \(^{235}\text{U}\) clusters, leading to a seven times increase in high-energy electrons.  This increase enables electron energies to excite the fourth nuclear state (81.724 keV), whose cascading decay efficiently accumulates the IS. This work provides a theoretical basis for optimizing laser parameters to achieve efficient generation of the IS, paving the way for experimental validation in laser-driven nuclear excitation schemes, and supporting applications in precision metrology and controlled nuclear energy.

\Acknowledgements{This work was supported by the National Natural Science Foundation of China (Grant No.12135009, 12375244), the Hunan Provincial Innovation Foundation for Postgraduate (Grant No.CX20230008), and the Innovation Foundation for Postgraduate (Grant No.XJQY2024046, XJJC2024063).}

\InterestConflict{The authors declare that they have no conflict of interest.}




\end{multicols}

\begin{thebibliography}{99}
\bibitem{MAIMAN1960} T. H. Maiman, Nature 187, 493 (1960).

\bibitem{STRICKLAND1985447} D. Strickland and G. Mourou, Opt. Commun. 55, 447 (1985).

\bibitem{Yoon21} J. W. Yoon et al., Optica 8, 630 (2021).

\bibitem{Xu2024} L. Xu and E. J. Takahashi, Nat. Photonics 1, 99 (2024).

\bibitem{doi10.1126/science.1080552} K. W. D. Ledingham et al., Science 300, 1107 (2003).



\bibitem{CPC-2021-0034} W. J. Huang et al., Chin. Phys. C 45, 030002 (2021).

\bibitem{CPC-2020-0033} M. Wang et al., Chin. Phys. C 45, 030003 (2021).

\bibitem{NUBASE2020} F. G. Kondev et al., Chin. Phys. C 45, 030001 (2021).


\bibitem{Gales_2018} S. Gales et al., Rep. Prog. Phys. 81, 094301 (2018).
\bibitem{10.1063/1.5093535} K. A. Tanaka et al., Matter Radiat. Extremes 5, 024402 (2020).




\bibitem{PhysRevLett.119.202501} D. S. Delion and S. A. Ghinescu, Phys. Rev. Lett. 119, 202501 (2017).

\bibitem{PhysRevLett.124.212505} A. P\'{a}lffy and S. V. Popruzhenko, Phys. Rev. Lett. 124, 212505 (2020).


\bibitem{PhysRevLett.133.152503} H. Zhang et al., Phys. Rev. Lett. 133, 152503 (2024).

\bibitem{PhysRevLett.133.132501} S. Gargiulo et al., Phys. Rev. Lett. 133, 132501 (2024).

\bibitem{PhysRevC.110.064326} T. Kirschbaum et al., Phys. Rev. C 110, 064326 (2024).

\bibitem{PhysRevLett.128.162501} Y. Wu et al., Phys. Rev. Lett. 128, 162501 (2022).

\bibitem{PhysRevC.103.014313} N. Minkov and A. P\'{a}lffy, Phys. Rev. C 103, 014313 (2021).

\bibitem{cjh2025} J. H. Cheng et al., Nucl. Sci. Tech. 36, 69 (2025).

\bibitem{10.1063/5.0212163} Z. Ma et al., Matter Radiat. Extremes 9, 055201 (2024).

\bibitem{PhysRevC.110.064621} Y. Y. Xu et al., Phys. Rev. C 110, 064621 (2024).

\bibitem{Xiao2024} Q. Xiao et al., Nucl. Sci. Tech. 35, 27 (2024).

\bibitem{Zou_2024} Y. T. Zou et al., J. Phys. G 51, 045103 (2024).

\bibitem{CHENG2024138322} J. H. Cheng et al., Phys. Lett. B 848, 138322 (2024).

\bibitem{PhysRevC.105.024312} J. H. Cheng et al., Phys. Rev. C 105, 024312 (2022).



\bibitem{PhysRevC.106.064604} B. Liu and X. Wang, Phys. Rev. C 106, 064604 (2022).

\bibitem{PhysRevC.106.044604} H. Zhang et al., Phys. Rev. C 106, 044604 (2022).

\bibitem{PhysRevLett.124.242501} E. V. Tkalya, Phys. Rev. Lett. 124, 242501 (2020).



\bibitem{PhysRevLett.128.052501} J. Feng et al., Phys. Rev. Lett. 128, 052501 (2022).

\bibitem{pnas2413221121} J. Feng et al., Proc. Natl. Acad. Sci. USA 121, e2413221121 (2024).

\bibitem{Shvydko2023} Y. Shvyd'ko et al., Nature 622, 471 (2023).

\bibitem{chiara2018isomer} C. J. Chiara et al., Nature 554, 216 (2018).

\bibitem{PhysRevLett.128.242502} S. Guo et al., Phys. Rev. Lett. 128, 242502 (2022).

\bibitem{Kraemer2023} S. Kraemer et al., Nature 617, 706 (2023).

\bibitem{PhysRevLett.132.182501} J. Tiedau et al., Phys. Rev. Lett. 132, 182501 (2024).

\bibitem{PhysRevLett.133.013201} R. Elwell et al., Phys. Rev. Lett. 133, 013201 (2024).

\bibitem{Zhang2024} C. K. Zhang et al., Nature 633, 63 (2024).

\bibitem{Zhang20242} C. K. Zhang et al., Nature 636, 603 (2024).

\bibitem{PARIKH2013225} A. Parikh et al., Prog. Part. Nucl. Phys. 69, 225 (2013).

\bibitem{Misch_2021} G. W. Misch et al., Astrophys. J. Suppl. S 252, 2 (2020).

\bibitem{10.3389/fphy.2024.1503516} B. Guo et al., Front. Phys. 12, 1503516 (2024).




\bibitem{PhysRevLett.128.043001} V. M. Shabaev et al., Phys. Rev. Lett. 128, 043001 (2022).

\bibitem{PhysRevResearch.7.L022036} F. Schaden et al., Phys. Rev. Res. 7, L022036 (2025).

\bibitem{PhysRevLett.134.113801} J. S. Higgins et al., Phys. Rev. Lett. 134, 113801 (2025).

\bibitem{PhysRevResearch.7.013052} S. V. Pineda et al., Phys. Rev. Res. 7, 013052 (2025).
\bibitem{BROWNE2014205} E. Browne and J. K. Tuli, Nucl. Data Sheets 122, 205 (2014).
\bibitem{PhysRevC.97.054310} F. Ponce et al., Phys. Rev. C 97, 054310 (2018).
\bibitem{PhysRevC.110.L051601} J. Qi et al., Phys. Rev. C 110, L051601 (2024).




\bibitem{SALVAT2019165} F. Salvat and J. M. Fern\'{a}ndez-Varea, Comput. Phys. Commun. 240, 165 (2019).

\bibitem{PhysRev.101.1131} B. Hahn et al., Phys. Rev. 101, 1131 (1956).

\bibitem{PhysRev.81.385} J. C. Slater, Phys. Rev. 81, 385 (1951).


\bibitem{PhysRev.99.510} R. Latter, Phys. Rev. 99, 510 (1955).

\bibitem{RevModPhys.30.353} K. Alder et al., Rev. Mod. Phys. 30, 353 (1958).

\bibitem{Arber_2015} T. D. Arber et al., Plasma Phys. Contr. F. 57, 113001 (2015).

\bibitem{PhysRevLett.121.253002} J. C. Berengut, Phys. Rev. Lett. 121, 253002 (2018).





\end{thebibliography}
\end{document}